# Design and Characterization of a Field-Switchable Nanomagnetic Atom Mirror

T.J. Hayward[1], A.D. West[2], K.J. Weatherill[2], P.J. Curran[3], P.W. Fry[4], P.M. Fundi[1], M.R.J. Gibbs[1], T. Schrefl[1], C.S. Adams[2], I.G. Hughes[2], S.J. Bending[3] and D.A. Allwood[1]

[1]*Department of Materials Science and Engineering, University of Sheffield, Sheffield, UK, S1 3JD*

[2]*Atomic and Molecular Physics Group, University of Durham, Durham, UK, DH1 3LE*

[3]*Nanoscience Group, University of Bath, Bath, UK, BA2 7AY*

[4]*Nanoscience and Technology Centre, University of Sheffield, Sheffield, UK, S3 7HQ*

Abstract

We present a design for a switchable nanomagnetic atom mirror formed by an array of 180° domain walls confined within $Ni_{80}Fe_{20}$ planar nanowires. A simple analytical model is developed which allows the magnetic field produced by the domain wall array to be calculated. This model is then used to optimize the geometry of the nanowires so as to maximize the reflectivity of the atom mirror. We then describe the fabrication of a nanowire array and characterize its magnetic behavior using magneto-optic Kerr effect magnetometry, scanning Hall probe microscopy and micromagnetic simulations, demonstrating how the mobility of the domain walls allow the atom mirror to be switched "on" and "off" in a manner which would be impossible for conventional designs. Finally, we model the reflection of $^{87}Rb$ atoms from the atom mirror's surface, showing that our design is well suited for investigating interactions between domain walls and cold atoms.

Introduction

Contemporary laser-cooling techniques [1] allow clouds of atoms to be cooled to temperatures within one millionth of a degree of absolute zero. These ultra-cold atoms are ideal systems with which to investigate quantum degenerate matter [2,3], and also have great technological potential through development of matter wave interferometry [4], atomic lithography [5,6], novel sensors [7], and in quantum information processing [8,9].

Common to all of these applications is a necessity to closely control the positions and motions of cold atoms. For example, there is great interest in developing "atom optics", devices which manipulate beams of cold atoms in a manner analogous to conventional optical components such as mirrors, beam splitters and diffraction gratings [10-12]. Other applications require that cold atoms are confined into potential wells, thus creating atom traps or waveguides [10-13].

Two main approaches can be used to manipulate atoms: Firstly, optical forces may be used, either through the optical dipole force as in an optical lattice [14] or an evanescent wave mirror [15,16] or via the scattering force, as in a magneto-optic trap (MOT) [17]. Alternatively, in the case of paramagnetic atoms, magnetic field gradients produced by either current carrying wires [18] or patterned ferromagnetic structures [19] may be used. Although optical techniques are currently more commonly used, magnetic manipulation has the potential to be highly important in future "atom chip" devices, due to the ease with which micrometer-scale current lines and ferromagnetic structures may be fabricated. Atom chips made from ferromagnetic films are particularly attractive due to their robustness and low power consumption.

To create atom traps/atom optics from a ferromagnetic material it must be patterned with a well-defined, spatially varying magnetization structure. Several techniques have been applied to achieve this. These have included writing magnetization patterns on audio and video tapes [20,21], creating stacks of macroscopic permanent magnets [22], lithographically patterning ferromagnetic films [23-25], optically writing magneto-optical films [26,27] and etching patterns into granular hard-disk media [28]. A common factor of all of these approaches is that they use hard magnetic materials, i.e. materials with strong magnetocrystalline anisotropies that dominate the behavior of their magnetization. Contrastingly, the behaviors of softer magnetic materials, such as $Ni_{80}Fe_{20}$, are dominated by magnetic shape anisotropy, meaning that their geometry directly controls their magnetization structure. This offers new routes to obtaining non-uniform magnetization structures which are suitable for manipulating atoms.

In a previous study [29] we have presented theoretical calculations which demonstrate the feasibility of creating zero dimensional atom traps using the magnetic fields emanating from 180° domain walls (DWs) in $Ni_{80}Fe_{20}$ nanowires [30]. These DWs are intrinsic to the nanowire geometry, and may be formed easily and reproducibly in positions governed by the shape of the nanowires using uniform external magnetic fields. This is in contrast to the intricate writing procedures required when using harder magnetic materials [20,21,26,27]. The DWs also have sub-micrometer dimensions and hence create high magnetic field gradients, thus offering extremely tight confinement of trapped atoms. Another important property is the reconfigurability of the DWs, which can be propagated controllably around nanowire circuits using applied magnetic fields or electric currents [31-33]. This offers the exciting possibility of mobile atom traps which could be used to transport qubits in quantum computing applications. The reconfigurability offered by DWs differs from that offered by arrays of current-carrying wires [34] in that nanomagnetic technology is inherently non-volatile, and hence power is only dissipated when a system's configuration is altered. The freedom with which individual

DWs may be moved around nanowire networks may also offer a wider range of magnetic field configurations than can be achieved using simple conductors.

In this paper we present the design and characterization of a simple nanomagnetic atom mirror which will be used to investigate experimentally the interaction of cold atoms with DWs in $Ni_{80}Fe_{20}$ nanowires. We first develop a model which allows the geometry of an effective reflecting surface to be calculated. We then go on to discuss the fabrication of the atom mirror and characterize its magnetic behavior, showing how the mobility of the DWs within the nanowires allows the mirror to be switched "on" and "off" in a manner which could not be achieved using conventional ferromagnetic designs. Finally, we model the dynamics of $^{87}Rb$ atoms reflecting from the surface of the atom mirror, thus demonstrating the feasibility of future experiments where these interactions will be probed experimentally. Experiments of this nature will lay a foundation for studies in which DWs will be used to trap cold atoms.

Theory

A paramagnetic atom moving adiabatically in a magnetic field gradient $\nabla |\mathbf{B}|$ will experience a force:

$$\mathbf{F} = -m_F g_F \mu_B |\nabla \mathbf{B}| \quad (1)$$

where $m_F$ is the atom's magnetic quantum number, $g_F$ is the Landé g-factor and $\mu_B$ is the Bohr magneton. If the atom is in a state where $m_F g_F > 0$ (i.e. its magnetic moment is quantized such that it has a component opposing the quantization axis) it is repelled by regions of increasing magnetic field, and hence is said to be in a "weak-field seeking" state.

Conventional magnetic mirrors are created using a sinusoidally varying magnetization pattern. This magnetization distribution produces lines of constant magnetic field which lie parallel to the film's surface [22]. The magnetic field decays exponentially, and hence it is a good approximation to assume that an atom in a weak-field seeking state incident upon the film with kinetic energy $U$ perceives a flat reflecting surface at height $z_r$ when:

$$U = m_F g_F \mu_B |\mathbf{B}(\mathbf{z}_r)| \quad (2)$$

Any higher harmonics present in the film's magnetization profile corrugate the reflecting surface [35]. This effectively makes the mirror's surface rough, such that the atom's dynamics contain an element of diffuse reflection.

To replicate such a mirror using DWs in magnetic nanowires we propose the undulating nanowire geometry shown in Figure 1(a). When a magnetic field is used to saturate and then relax these nanowires perpendicular to their length ($H_y$) head-to-head (H2H) and tail-to-tail (T2T) DWs will be formed at the apexes of the curves ("on" state) . The DWs have a dominant monopolar character (North or South), and hence the nanowires can be considered as an array of magnetic poles with alternating polarity. This is a discretized equivalent of the pole distribution in conventional magnetic mirrors, where the sinusoidal magnetization results in a cosinusoidal pole distribution.

Unlike magnetic mirror designs which utilize hard magnetic materials, the design we propose allows the atom mirror's magnetization to be reconfigured such that its reflecting properties are "switched off". Shape anisotropy favors magnetization that is continuous along the nanowires, and hence by applying a magnetic field along $H_x$ a state similar to that shown in Figure 1(b) will be formed ("off" state). As this state contains no DWs, no magnetic field is created above the nanowire array.

To model the magnetic fields produced by the "on" state we approximate each DW as a point-monopole, as shown schematically in Figure 1(c). The array in the figure is curtailed for clarity; much larger arrays were used in the calculations presented here (~$10^4$ DWs). At a position **r** above the array the magnetic scalar potential is calculated using:

$$\phi(\mathbf{r}) = \sum_{array} \frac{q_i}{4\pi\mu_0 |\mathbf{r} - \mathbf{r_i}|} \qquad (3)$$

Where $q_i$ is the total "magnetic charge" of each of the DWs and $\mu_0$ is the permeability of free space. The magnetic field can then be calculated using:

$$\mathbf{H}(\mathbf{r}) = -\nabla \phi(\mathbf{r}) \qquad (4)$$

$q_i$ may be calculated by evaluating the following intergral over the volume of a DW:

$$q_i = -\mu_0 \iiint_V \nabla \cdot \mathbf{M} \, dV \qquad (5)$$

Because the total magnetic charge of a H2H or T2T DW, $q_i$, is independent of its internal magnetization structure $q_i$ can be derived analytically by considering the abrupt DW shown in Figure 1(d). The integral is then easily evaluated by invoking the divergence theorem and summing the magnetization flux entering the highlighted region:

$$q_i = \pm 2\mu_0 M_s wt \qquad (6)$$

A positive charge represents a H2H wall (North monopole), while a negative charge represents a T2T wall (South monopole). $w$ and $t$ are the nanowire width and thickness, and $M_s$ = 860 kA/m is the saturation magnetization of $Ni_{80}Fe_{20}$. As is shown in Figure 1(c), end domains must be assigned charges half this magnitude to ensure that the array carries no net monopole moment.

Modeling

To optimize the geometry of the nanowire array for use as an atom mirror we consider a simple experiment in which a cloud of atoms is "bounced" from the mirror under gravity. For $^{87}$Rb atoms in the $5^2S_{1/2}$ $F = 2$, $m_F = 2$ state, dropped onto the mirror from a height of 1 cm, $|\mathbf{H}_{reflect}| = 15.7$ Oe will be required to reflect the incident atoms (Equation 2).

We performed calculations to understand how the nanowire geometry affects the total "reflectivity" of the array, i.e. the proportion of the array's surface above which a magnetic field in excess of $|\mathbf{H}_{reflect}|$ is obtained. The calculations were performed for a 100 x 100 cell discretization of the central unit cell in a 100 x 100 μm$^2$ array of nanowires. Two parameters were varied: the thickness of the nanowires, $t$, and the average diameter of nanowire curvature $d$ (indicated in Figure 1(c)). To maximize $q_i$, while retaining wire-like magnetic properties, we set the width of the nanowires, $w = 0.25d$.

Figure 2(a) illustrates how the mirror's reflectivity varies with $d$ for constant $t = 30$ nm. At large $d$ the $|\mathbf{H}_{reflect}|$ field surface is composed of approximately spherical shells surrounding each of the DWs. As $d$ decreases these begin to cover a larger proportion of the array's surface, resulting in the rapid increase in reflectivity. However at low $d$ the surface is no longer composed of discrete shells, but is continuous from one DW to another and thus further decreasing $d$ has a reduced effect.

Figure 2(b) shows how the reflectivity varies with $t$ for a constant $d = 1$ μm ($w = 0.25$ μm). The plateau in the data is once again due to a transition from a field surface composed of discrete shells to a continuous surface.

Bearing in mind these results, we selected values of $d = 0.5$ μm and $t = 30$ nm for our experimental nanowire array. We limited $d$ to this value as further decreases would necessitate the fabrication of nanowires with sub-100 nm dimensions, which is challenging for the large (mm$^2$) arrays required experimentally. We limited $t$ to 30 nm to reduce the likelihood of forming complex DW structures [36] and to limit the applied field required to switch the array into the "on" state [37].

Two-dimensional (2D) micromagnetic simulations were performed with the OOMMF software package [38]. Simulated magnetization configurations of the "on" and "off" states are shown in Figures 3(a) and 3(b), respectively. A cell size of 5 x 5 nm$^2$ was used, while standard parameters were used to represent the material parameters of Ni$_{80}$Fe$_{20}$ (saturation magnetization, $M_S = 860$ kA/m, exchange stiffness A = 13 pJ/m, magnetocrystalline anisotropy constant $K_1 = 0$).

Both "on" and "off" states were found to be stable at remanence. However, the "on" state has a much higher energy ($E_{total} = 0.320$ fJ) than the "off" state ($E_{total} = 0.077$ fJ). This is primarily due to the large magnetostatic energy of the "on" state, resulting from the numerous DWs it contains. Consequently, in dynamical simulations, a much larger field was required to switch the nanowire from the "off" state into the "on" state ($H_y = 1000 \pm 20$ Oe) than was required to make the reverse transition ($H_x = 60 \pm 20$ Oe). These results suggest that the "off" state is the nanowire's magnetic ground state, while the "on" state is a meta-stable configuration which is stabilized by the localized pinning of DWs. A consequence of this is that localized defects are likely to enhance the reflectivity of the "on" state by ensuring that the DWs are well pinned, such that the nanowires do not collapse spontaneously into the lower energy "off" state. In the simulations DW pinning is induced by edge roughness resulting from the discretization of the nanowires into a regular mesh, while in a real nanowire array both lithographic and material defects would contribute.

The "on" to "off" transition was found to proceed via the annihilation of adjacent H2H and T2T DWs (Figure 3(c)), while the "off" to "on" transition occurred via the nucleation of DW pairs in the sections of the nanowire where the magnetization was anti-parallel to the applied field (Figure 3(d)). This behavior is analogous to that of micrometer-scale ferromagnetic rings [39], structures to which they are geometrically similar. The "on" state is equivalent to the "onion" state in the ring geometry, while the "off" state is equivalent to the "vortex" state. It is possible that in a real nanowire array DW pinning by random defects would increase the fields required to switch between "on" and "off" states. However, it is highly unlikely that the fields required to obtain a perfect "off" state would exceed those which are required to switch the mirror "on", and hence the presence of defects will have a negligible effect on the feasibility of reconfiguring the mirror experimentally.

Experimental Technique

2 x 2 mm² arrays of nanowires were patterned on Si/SiO$_2$ substrates using electron-beam lithography with lift-off processing. To facilitate the fabrication of such large arrays, which contained ~2000 individual 2 mm long nanowires, each nanowire was written into poly(methyl methacrylate) (PMMA) resist as a single pixel line using an electron dose calibrated to give the desired 125 nm width. Metallization with Ni$_{80}$Fe$_{20}$ was achieved by thermal evaporation with a base pressure ~$10^{-7}$ mbar. Scanning electron microscopy (SEM) images of the fabricated nanowire array are presented in Figure 4(a).

Magnetic hysteresis measurements of the nanowire arrays were performed using the Magneto-Optical Kerr Effect (MOKE). Measurements were performed in the transverse geometry with applied magnetic fields up to 2 kOe.

Quantitative imaging of the magnetic fields produced by the nanowire arrays was achieved by scanning Hall probe (SHP) microscopy at a temperature of 88 K. The Hall probe was aligned to measure the out-of-plane component of the magnetic field ($H_z$), and had an active area of ~ 200 x 200 nm, which limited the spatial resolution of the measurements. The measurements were performed with scan heights in excess of 700 nm to avoid damaging the probe through contact with dust particles and residual resist on the sample's surface.

Magnetic Characterization

Figure 4(b) presents a MOKE loop measured from the nanowire array with both applied field and magnetization sensitivity along $H_y$. The array exhibits a double-stepped switching behavior indicating that at least two distinct configurations were formed during its reversal.

The high-magnetization remanent state, which is formed reversibly from positive or negative saturation, is the "on" state. We term the second state, which is formed following the application of a small (< 100 Oe) reverse field the "quasi-off" ("q-off") state. As we will show later, this state is similar to the "off" state in that the nanowires' magnetization is predominantly continuous; however, here each nanowire contains a few large domains, and consequently a small number of DWs. The "q-off" state is formed rather than the "off" state in this measurement because a uniaxial field sequence is used to induce the array's switching, rather than the biaxial sequence described earlier. Domains pointing along the nanowires in directions +x and –x are thus degenerate, unlike in the biaxial case where the application of a field along $H_x$ creates a unidirectional anisotropy. Schematic diagrams of the "on" and "q-off" states are shown in Figure 4(b).

The micromagnetic simulations presented earlier indicate that the "on" state is metastable. Therefore it is likely that some of the less strongly pinned DW pairs will annihilate spontaneously, reducing the atom mirror's reflectivity. To estimate the percentage of DW pairs present in the nanowire array's remanent state we differentiated the "on" to "q-off" transitions and fitted Gaussian distributions to the resulting data (see inset to Figure 4(b)). Integrating these distributions then allowed the number of DW present at a given applied field to be calculated. Using this method we estimate that ~98 % of DW pairs are stable at remanence. Spontaneous DW annihilation is therefore likely to have only a small effect on the array's total reflectivity. However, it should be noted that over timescales longer than those of this measurement (~minutes) thermal excitation may cause further DW pairs to de-pin and annihilate. Therefore if the mirror is to be used for extended periods of time it may be necessary to occasionally "reinforce" its magnetic configuration by reapplying an external field.

In Figure 5 we present SHP microscopy images which illustrate the switching of the nanowire array using fields along both $H_y$ (Figure 5(a)) and $H_x$ (Figure 5(b)). In both cases the array was switched initially into the "on" state by applying briefly $H_y \sim$ -1 kOe. Consequently, Figure 5(a) corresponds to the application of a uniaxial field sequence (as in the MOKE data c.f. Figure 4b), while Figure 5(b) corresponds to a biaxial field sequence (as in the micromagnetic simulations c.f. Figure 3).

The SHP images of the nanowire array's remanent state exhibit a checkerboard pattern of light and dark regions, each of which corresponds to the position of a DW. Above H2H DWs $H_z$ is positive, resulting in light contrast, while above T2T DWs $H_z$ is negative, resulting in dark contrast. The checkerboard patterns are unbroken in the absence of applied field indicating that all DW sites in the imaged area are filled.

Figure 5(a) illustrates how the SHP images evolved in response to a uniaxial field sequence. As the applied field is increased the regular checkerboard pattern becomes progressively disordered due to the pair-wise annihilation of some DWs, and the isolated motion of others. In the final image, taken with $H_y$ = 360 Oe, there are a small number of well-separated DWs. The nanowires' magnetization is therefore dominantly continuous, but is multi-domain over longer length-scales, as in the "q-off" state described earlier. Increasing the applied field to $H_y \sim 750$ Oe produced no further DW annihilation. This is to be expected for the "q-off" state, where any remaining DWs are isolated from DWs with opposite H2H or T2T character.

Figure 5(b) illustrates how the nanowire array's switching behavior was altered when a biaxial field sequence was applied. As the field increases the checkerboard pattern once again becomes disordered. However, in contrast to the uniaxial switching images, DWs are only ever present in pairs, and no isolated DWs are observed in any of the images. At $H_x$ = 91 Oe only two DW pairs are present, and the nanowires' magnetization is dominantly continuous and single domain. Increasing $H_x$ further caused no additional changes in the images, most likely because the measurement was performed at 88 K, resulting in enhanced DW pinning over that at room temperature. However, had larger fields been available and/or the measurements been performed at room temperature it is likely that the remaining DW pairs would have annihilated, forming a true "off" state. The SHP data therefore illustrates the importance of using a biaxial, rather than a uniaxial, field sequence to switch the atom mirror between the "on" and "off" states.

In Figure 6(a) we present a quantitative comparison between a SHP microscopy image and the predictions of the point-monopole model. As the scan height of the Hall probe was not known it was treated as a free parameter and was optimized to produce good agreement with the measured values of $H_z$. Using this method we calculated a scan height **z** = 780 nm. The finite size of the Hall probe was taken into account at each point by averaging the predicted fields over a 200 nm x 200 nm area.

It can be seen that there is good agreement between the predictions of the point-monopole model and the experimental data. This is illustrated further in Figure 6(b) in which we compare values of $H_z$ measured along the dashed lines in Figure 6(a). The differences between the two curves are caused by a slight disordering of the positions of the DWs in the SHP images. This may be attributed to a random distribution of lithographic and material defects within the nanowires.

The data shown in Figure 6 indicate that for $z \sim 800$ nm the point monopole model provides an excellent approximation of the nanowire array's magnetic field. However, this represents a far-field case and in the experiment we propose atoms will approach much closer to the array's surface. At these heights it is expected that the finite size and magnetization structure of the DWs will play a more important role in determining the magnetic field, and hence the point-monopole model will not be such a good approximation. We will discuss these effects further in the following section.

Simulated Reflection Dynamics

In this section we will calculate the behavior of a cloud of cold $^{87}$Rb atoms reflecting from the atom mirror, thus demonstrating the feasibility of using our design in initial investigations into the interaction of cold atoms with DWs. However, before presenting these simulations we will first present calculations which justify our methodology.

As we showed above, the point-monopole model is an excellent far-field approximation of the magnetic field created by the nanowire array. However, for $^{87}$Rb atoms in the $5^2S_{1/2}$ $F = 2$, $m_F = 2$ state, dropped from $z = 1$ cm directly onto a DW, the same model predicts that reflection will occur at $z \sim 450$ nm where the finite size and magnetization structure of the DWs is likely to be important.

To investigate this we calculated the magnetic field pattern above a single DW using a micromagnetic modeling package that uses a hybrid finite element/boundary element technique [40]. The parameters used to represent $Ni_{80}Fe_{20}$ were the same as those used in the OOMMF simulations except for a slightly lower value of $M_s = 796$ kA/m.

In these simulations we found that the DW had vortex structure (see inset Figure 7(a)) rather than transverse structure as had been seen in the OOMMF simulations. We believe that this is in fact a better representation of the DWs in the experimental nanowires, as evidenced by the small steps proceeding to the "on" to "q-off" transitions in the MOKE loop shown in Figure 4(b), which most likely represents the relaxation of a metastable transverse DW configuration into a lower energy vortex configuration. The transverse DW configurations in the OOMMF simulations are likely to be stabilized by the effective edge roughness of the rectangular finite difference mesh.

In Figure 7(a) we present simulated values of $H_z$ as a function of $z$, directly above the centre of the DW. Also shown is a fit through the data using the point-monopole model with the magnetic charge, *q*, as a free parameter. It can be seen that there is excellent agreement between the simulated and point monopole data. However the fitted value of *q* is 9 % lower than that calculated using Equation 6. We believe that this reflects the fact that at these heights the finite size of the DW's magnetic charge distribution reduces the gradient of the magnetic scalar potential with respect to the point-monopole model. Nevertheless, the data indicate that the point-monopole model is a useful approximation, and is suitable for use in this feasibility study. In an attempt to compensate for the effects of finite DW size we use the reduced value of $q$ in the following calculations.

In conventional magnetic mirror designs the magnetic field above the mirror's surface decays as $|\mathbf{H}| \propto e^{-\left(\frac{2\pi z}{\lambda}\right)}$ where $\lambda$ is the wavelength of the mirror's periodic magnetization configuration [22]. Consequently an atom's interaction time with the mirror is extremely short and an atom's dynamics can be modeled as an instantaneous, specular reflection from the iso-field surface at which $|\mathbf{H}| = |\mathbf{H}_{reflect}|$. We will now demonstrate that a similar approach is well suited to the atom mirror design we propose.

In Figure 7(b) we present a log-linear plot of $|\mathbf{H}|$ as a function of $\mathbf{z}$ as calculated for our atom mirror design using the point-monopole model. Data are shown for positions ( x μm, 50 μm, z ), where x = 50 μm → ( 50 + 0.8$d$ ) μm (see inset figure). We choose to evaluate data along this line because at x = ( 50 + $d$ ) μm the symmetry of the magnetic charge distribution indicates that $|\mathbf{H}(\mathbf{z})| = 0$, and therefore we might expect to see "softening" of the magnetic field gradients as we approach this point.

For all values of x the plots retain a high degree of linearity up to $|\mathbf{H}_{reflect}|$. Fitting in this region produces $\lambda = 0.950 \pm 0.006$ μm, close to the array's actual 1 μm wavelength. The small deviation from the expected value is most likely due to the presence of higher harmonics in the mirror's magnetization configuration, as will be explored in more detail later. Because $|\mathbf{H}|$ decays exponentially as in conventional magnetic mirror designs, modeling an atom's dynamics as an instantaneous reflection from the $|\mathbf{H}_{reflect}|$ iso-field surface is a valid approach for the system we consider.

In Figure 8(a) we present the geometry of the $|\mathbf{H}_{reflect}|$ iso-field surface calculated using the point-monopole model. Data are shown for a 5 μm x 5 μm region close to the centre of the nanowire array. A more detailed plot of the array's central unit cell is also shown as an inset to Figure 8(b). The nominally spherical iso-field surfaces surrounding each of the DWs have merged together to form a continuous surface, however, a significant amount of curvature is still evident, such that the atom mirror's surface has a finite "roughness". This is illustrated further in Figure 8(b), which presents the distribution of angles the field surface makes to the horizontal within the unit cell. Experimentally, the mirror's roughness will mean that an atom cloud will be diffusely reflected. The field surface is also "porous", containing non-reflecting regions centered on the symmetry points described earlier, however we calculate that 97 % of the surface area of the unit cell is covered by fields in excess of $|\mathbf{H}_{reflect}|$, and therefore the porosity in itself should only have a minor effect on the reflectivity of the mirror.

In Figure 9 we present calculated reflection dynamics for $^{87}$Rb atoms dropped onto the array's unit cell at normal incidence from a height of 1 cm. The atoms are modeled at 0 K so as to isolate their intrinsic dispersion from that due to thermal effects. As would be expected from the histogram shown in Figure 8(b), the atoms are reflected diffusely with the greatest number of atoms having trajectories at ~ 40° to the vertical.

To assess the suitability of measuring the atom's reflection experimentally we define an "experimentally accessible" region into which a laser light sheet could be inserted to detect the atom's motion by resonant absorption. We define this region as being 2.5-7.5 mm above the surface of the nanowire array. It can be seen that the majority of the reflected atoms pass through the experimentally accessible region within 20 ms of leaving the surface of the mirror. This result demonstrates the feasibility of detecting the atom's motion experimentally, and indicates that the atom mirror we propose is suitable for investigating interactions between domain walls and cold atoms.

It is also likely that our simple theoretical treatment overestimates the roughness of the $|\mathbf{H}_{reflect}|$ field surface. In the point-monopole model the nanowires' magnetic charge distribution is described by a 2D array of delta functions in both real and Fourier space. The model's charge distribution therefore contains a vast number of high frequency harmonics which contribute to the atom mirrors roughness. In reality the DWs have finite widths and hence higher frequency components will be less significant.

The effect of this will be a smoothing of the $|\mathbf{H}_{reflect}|$ field surface in comparison to that predicted by the point-monopole model.

Conclusions

In this paper we have described the design, fabrication and characterization of a reconfigurable nanomagnetic atom mirror which will be used in initial investigations into the interactions between DWs and cold atoms. The atom mirror is formed by the magnetic field emanating from a large number of 180° DWs confined within a periodic array of undulating planar magnetic nanowires.

We have demonstrated the fabrication of a nanowire array which will be large enough to be used to experimentally investigate the "bouncing" of an atom cloud upon the mirror under gravity. Magnetic characterization of the mirror indicates that ~98 % of potential DW sites in the array are filled at remanence, and also allowed us to show how the mobility of the DWs permits the atom mirror to be switched "on" and "off" using biaxial magnetic field sequences. We have also modeled the dynamics of $^{87}$Rb atoms reflecting from the atom mirror's surface. Our results indicate that although the mirror's surface exhibits substantial roughness, the atoms' reflections will be sufficiently well defined to allow experimental probing of their dynamics.

Cumulatively, the results of this study indicate that the atom mirror design we propose will be an ideal system with which to perform initial investigations into manipulating cold atoms using DWs in ferromagnetic nanowires. In addition to providing a system that will allow interactions to be observed over length-scales that are easily accessible experimentally, it will also allow us to demonstrate the reconfigurability intrinsic to atom optics/atom traps formed by magnetic nanostructures composed of soft ferromagnetic materials.

Future studies will focus on an experimental realization of the atom bouncing experiment we have modeled in this paper. This in turn will inform experiments in which we will attempt to achieve our ultimate goal of trapping a single atom in the potential well created by a DW. Nanowire arrays with geometries similar to that which we have investigated in this paper could ultimately be used to create two dimensional arrays of magnetic traps, which have been shown to be useful tools for studying phenomena such as Josephson oscillations [41] and Mott insulator transitions [42] in atomic ensembles.

The authors thank the Engineering and Physical Sciences Research Council for financial support (grant Nos. EP/F024886/1 and EP/F025459/1).

**Figure Captions**

Figure 1: Schematic diagrams indicating the nanowire array's magnetization configuration in (a) the "on" state and (b) the "off" state. (c) Illustration of how the distribution of magnetic charge in the "on" state is approximated in the point-monopole description of the nanowire array. (d) Schematic diagram of a head-to-head domain wall.

Figure 2: (a) Calculated reflectivity of the atom mirror as a function of $d$, the nanowires' diameter of curvature, for fixed thickness and width, $w = 0.25d$. (b) Calculated reflectivity of the atom mirror as a function of $t$, the magnetic film thickness, for constant width and diameter of curvature.

Figure 3: Micromagnetic simulations of a nanowire in (a) the "on" state and (b) the "off" state. The applied fields required to switch the nanowire between the "on" and "off" states are also indicated. (c) and (d) present the results of dynamical simulations and show the nanowires' behavior during "on" to "off" and "off" to "on" switching respectively. The portion of the nanowire shown in (c) and (d) is highlighted in (b).

Figure 4: (a) Scanning electron micrograph of the nanowire array. (b) MOKE loop of the nanowire array. The inset figure shows the differential of the data in the region of the "on" to "q-off" transition for the negative field sweep. The solid line shows the results of fitting a Gaussian distribution to the data. Schematic diagrams of the "on" and "q-off" state are also shown.

Figure 5: Scanning Hall probe microscopy images of the nanowire array taken as magnetic fields are applied along (a) +$H_y$ (uniaxial switching) and (b) +$H_x$ (biaxial switching). In both cases the nanowire array was initialized into the "on" state by the application of $H_y$ = -1 kOe.

Figure 6: (a) SHP image of the nanowire array in the "on" state and equivalent image generated using the point monopole model. (b) Data showing values of $H_z$ measured along the dashed lines in the SHP image (open circles) and point-monopole model data (continuous line).

Figure 7: (a) Results of a micromagnetic simulation in which $|\mathbf{H}|$ was measured as a function of the height, z, above an isolated domain wall confined in a straight nanowire with width and thickness identical to the nanowires measured experimentally. The continuous line is a fit to the data using a point-monopole approximation with the domain wall charge, $q_i$, as a free parameter. The inset figure shows the magnetization structure of the simulated domain wall. (b) Results of calculations performed using the point-monopole in which $|\mathbf{H}|$ was calculated as a function of z above various points in a 100 x 100 μm² undulating nanowire array. Data are shown for positions ( x μm, 50 μm, z ), where x = 50 μm (closed squares), 50 + 0.2d μm (closed circles), 50 + 0.4d μm (closed triangles), 50 + 0.6d μm (open squares) and 50 + 0.8d μm (open circles).

Figure 8: (a) Geometry of the $|\mathbf{H}_{reflect}|$ iso-field surface calculated using the point-monopole model. (b) Distribution of angles of the $|\mathbf{H}_{reflect}|$ iso-field surface makes to the horizontal within the central unit cell of atom mirror. A detailed plot of the $|\mathbf{H}_{reflect}|$ iso-field surface in the unit cell is shown inset.

Figure 9: Calculated dynamics of $^{87}$Rb atoms reflecting from the central unit cell atom mirror. The atoms are positions are shown as a function of their height above the mirrors surface, z, and their radial displacement in the plane of the substrate, r. The density of the reflected atoms is indicated by grayscale contrast. Data are shown (a) 5 ms, (b) 10 ms, (c) 15 ms and (d) 20 ms after the atoms leave the surface of the mirror. The dashed lines indicate an "experimentally accessible" region into which a light sheet could be inserted to experimentally measure the atoms' dynamics.

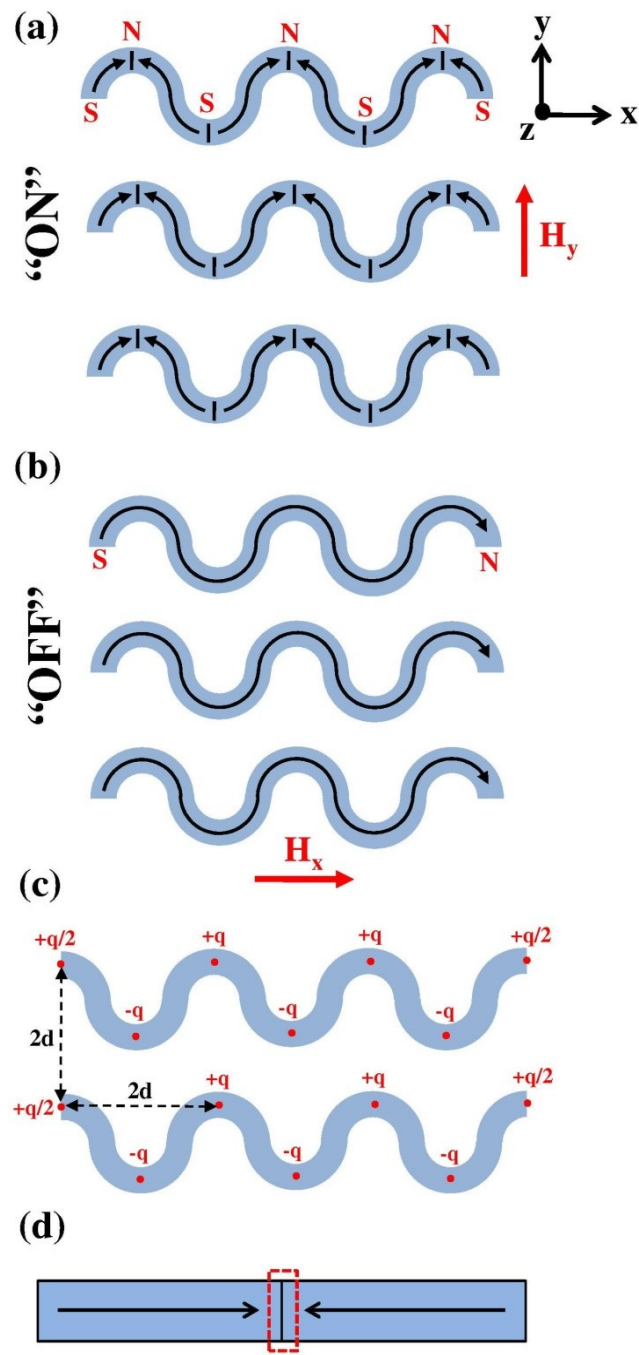

**Figure 1**

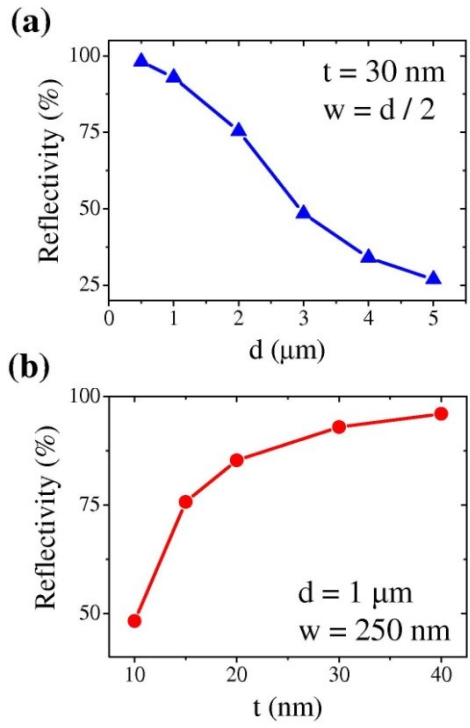

**Figure 2**

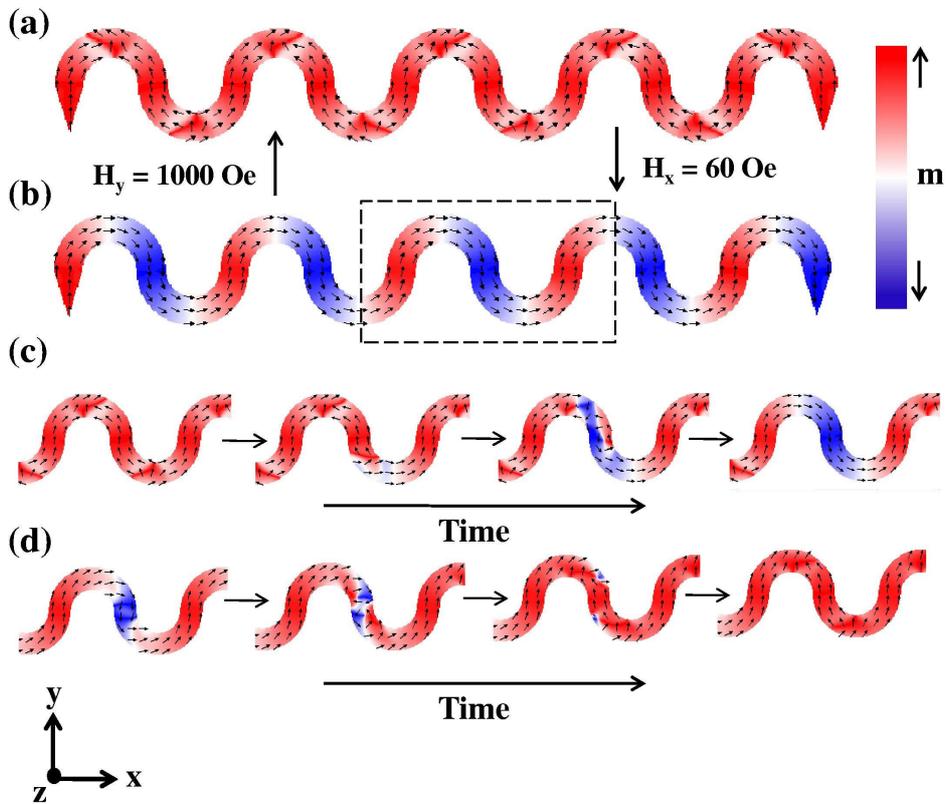

**Figure 3**

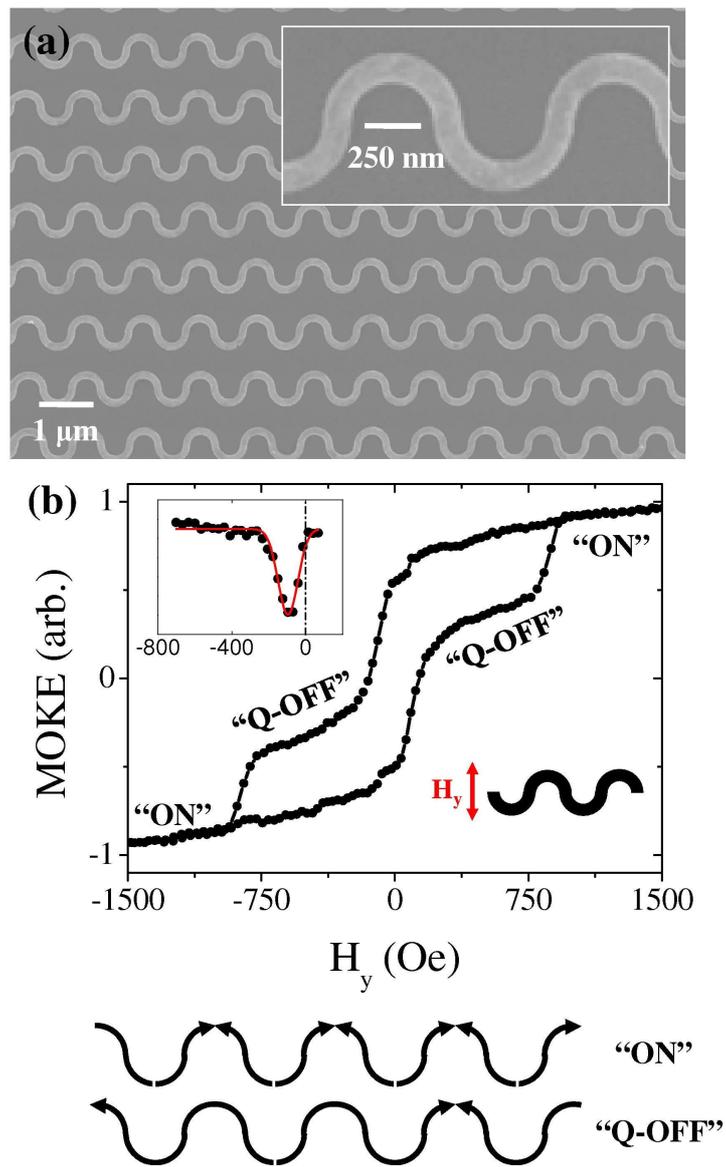

**Figure 4**

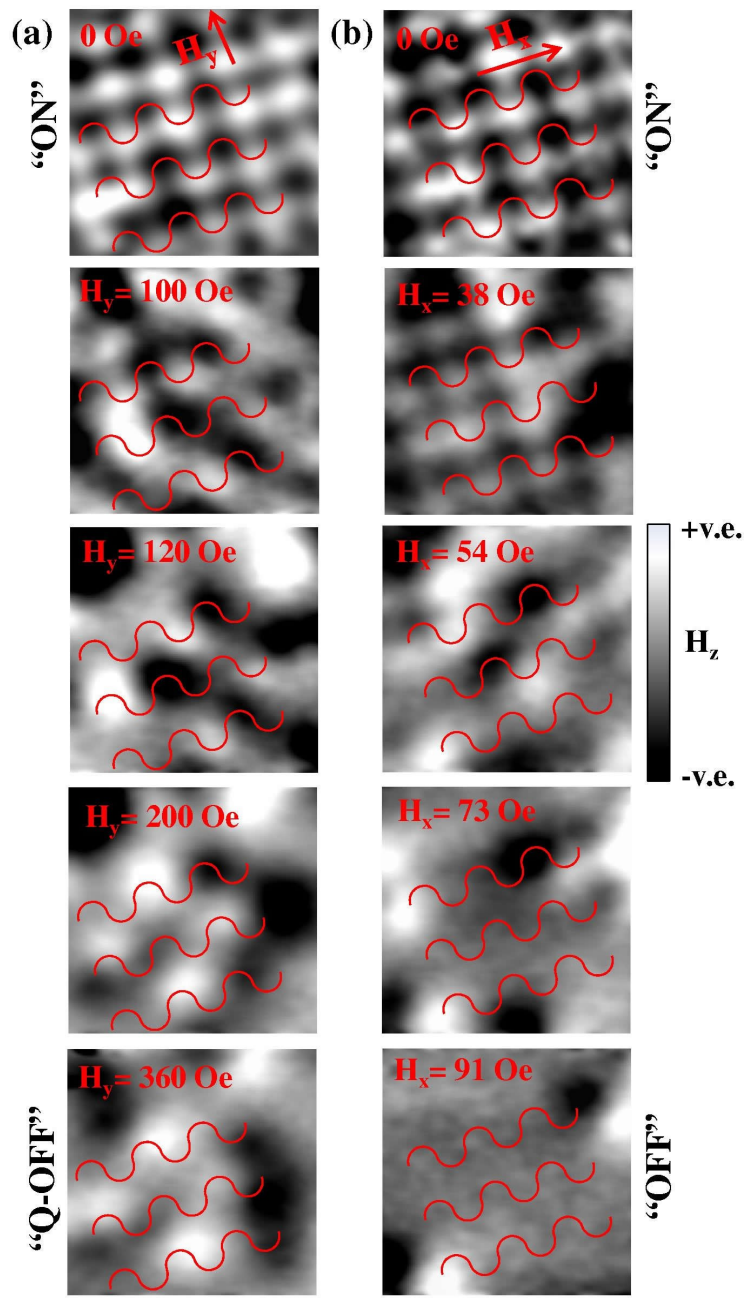

**Figure 5**

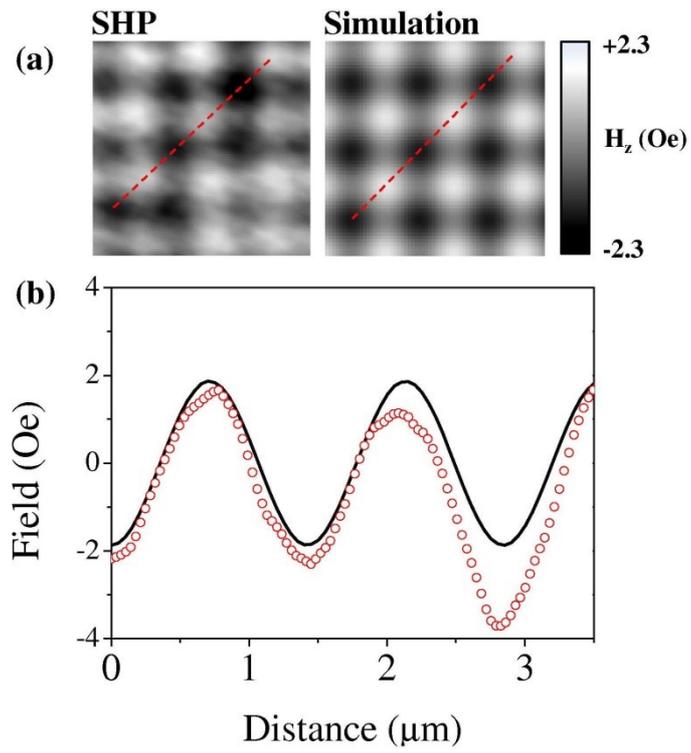

Figure 6

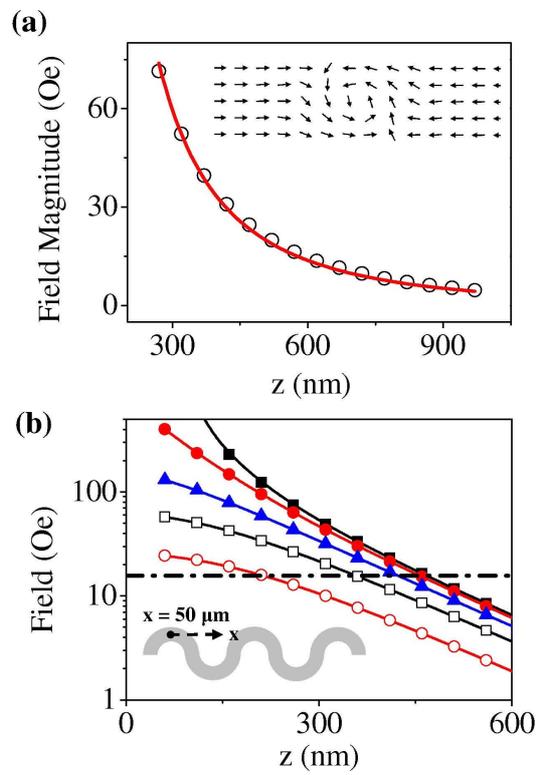

Figure 7

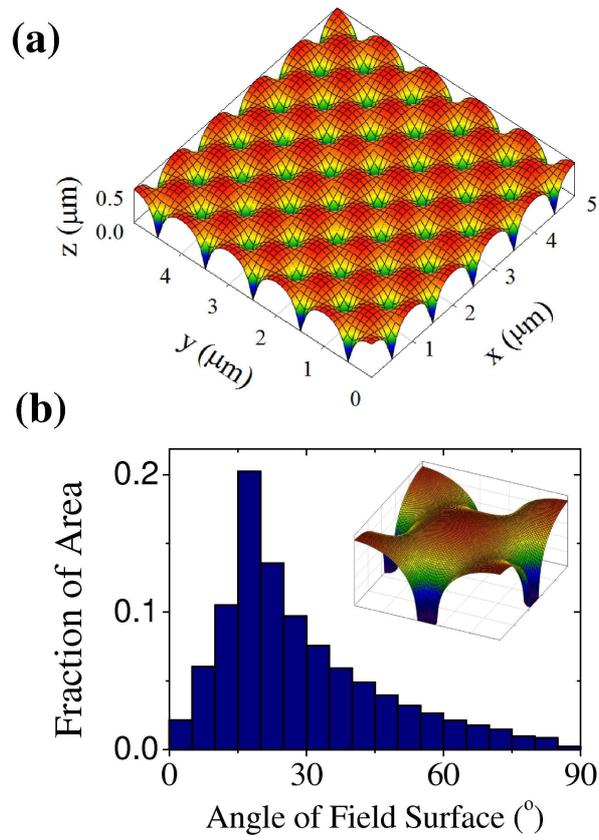

Figure 8

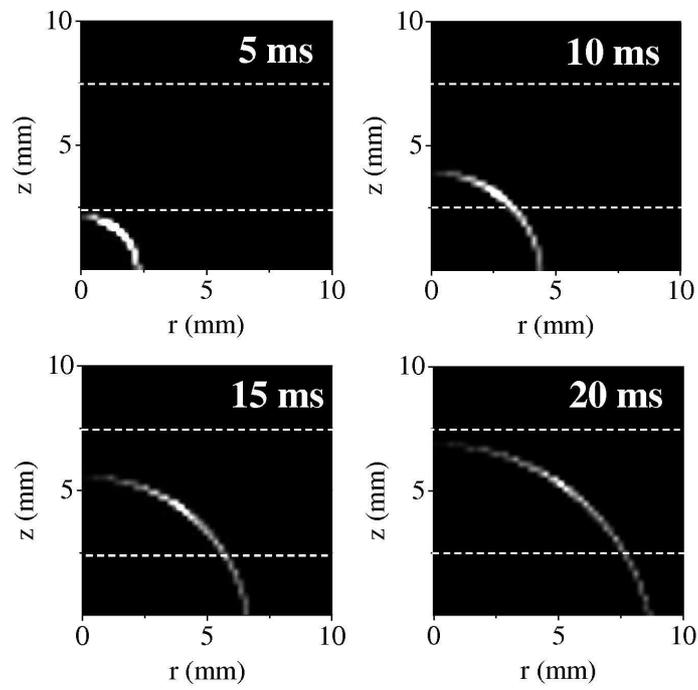

Figure 9